\documentclass{aa}
\usepackage{psfig}
\begin{document}
%
%\thesaurus{(02.01.2; 02.13.1; 08.14.1; 08.13.1; 13.25.5)}

   \title{Period distribution of old accreting isolated neutron stars}
%\subtitle{}

\titlerunning{Periods of accreting isolated neutron stars}

 \author{M.E. Prokhorov
          \inst{1}
          \and
          S.B. Popov\inst{2}
          \and
          A.V. Khoperskov\inst{3}
          }

   \offprints{S. Popov}

   \institute{Sternberg Astronomical Institute,
              Universitetski pr. 13, 119899 Moscow\\
%              \email{mystery@sai.msu.ru}
         \and
                Sternberg Astronomical Institute,
              Universitetski pr. 13, 119899 Moscow\\
%              \email{polar@sai.msu.ru}
          \and
             Volgograd State University,
              Department of Theoretical Physics,
               40068 Volgograd\\
%               \email{khoperskov@vlink.ru}
             }

   \date{}

   \abstract{
 In this paper we present calculations of period distribution for
old accreting isolated neutron stars (INSs). 
After few billion years of evolution
low velocity INSs come to the stage of accretion. At this stage INS's
period evolution is governed by magnetic braking and
angular momentum accreted.
Since the interstellar medium is turbulized accreted momentum
can either accelerate or decelerate spin of an INS, 
therefore the evolution of period has chaotic character.
 Our calculations show that in the case of
constant magnetic field accreting INSs
have relatively long spin periods (some hours and more,
depending on INS's spatial velocity, its magnetic field and
density of the surrounding medium). Such periods are much longer than the
values measured by {\emph ROSAT} for 3 radio-silent isolated neutron stars.
Due to long periods INSs should have high spin up/down rates, $\dot p$,
which should fluctuate on a time scale of $\sim 1$ yr.
}

\maketitle

\section{Introduction}

 Spin period is one of the most precisely determined
parameters of a neutron star (NS). 
Estimates of masses (for isolated objects),
radii, temperatures, magnetic fields etc. are usually
model dependent. 
That is why it is very important to have a good
picture of  evolution of the only model independent physical parameters
of INSs --- spin period and its derivative,  as they are 
usually used to determine other characteristics of NSs.
Our aim is to obtain distribution of periods $p$
for old accreting INSs (AINSs), we also briefly discuss 
possible values of period derivative $\dot p$.

 A lot has been done to understand period evolution of
radio pulsars (see \cite{bgi93}) and NSs in close binaries 
(\cite{gl79}, \cite{li92}). 
AINS are especially interesting from the point of view of period
(and by the way magnetic field) evolution since their history
is not ``polluted'' by huge accretion, as it happens with their relatives
in close binary systems, where NSs can accrete up to $1\, M_{\odot}$
during extensive mass transfer. 

In early 90$^{\rm s}$ there has been a great enthusiasm 
about the possibility to observe a huge
population of AINSs with \emph{ROSAT} (\cite{tc91}, \cite{br91},
\cite{bm93}, \cite{m94}). However, it has become clear that AINSs are very 
{\it elusive} (\cite{tetal98}) due to high spatial velocities
(\cite{pea00a}) or/and magnetic field properties (\cite{elu98}, \cite{lxf98}).

Still,  AINSs (and radio-silent INSs in general) 
are a subject of interest in astrophysics
(see \cite{car96}, \cite{treves00} and references therein).
A few candidates seem to be observed by \emph{ROSAT} (\cite{motch2000}),
although it is also  possible that
these sources (at least a part of them) 
can be better explained by young cooling NSs
(see \cite{nt1999}, \cite{yak99}, \cite{w00}, \cite{pea00b}).
Nevertheless, such sources should be very abundant at low fluxes
available for \emph{Chandra} and \emph{XMM--Newton} 
observatories (in \cite{pea00b} the authors obtain that at 
$\sim 10^{-13}$ erg s$^{-1}$ AINSs become more abundant than young cooling
NSs, and expected number is about 1 source per square degree for fluxes
$>10^{-16}$ erg s$^{-1}$), 
so the calculation of properties of AINSs is of great importance now.

 Previous estimates of spin properties of AINSs 
(\cite{lp95}, \cite{kp97}) have
given only typical values
of periods, no realistic distributions 
of this parameter are calculated.
``Spin equilibrium'' of NSs with the interstellar medium (ISM) has been
always assumed, i.e. the authors have considered the situation, when all INS
have enough time for spin evolution, they have not considered NSs with
relatively high spatial velocities. In this paper we present full analysis
of the problem. 

 Period estimates are especially important as far as this parameter can be
used to distingush accreting INSs from young cooling NSs and background
objects.

 We proceed as follows:
 In the next section we describe the model used to calculate period
distribution. In Section 3 we present our results. In this paper
we do not address the question of the total number of AINS, for this data
we refer to our previous calculations 
(\cite{pea00a}, \cite{pea00b}). Here we only
show period distributions for AINSs. In the last section
we give a brief discussion, derive typical parameters of $\dot p$ for AINSs 
and summarize the paper.

\section{Model}

 In this section we describe our model of spin evolution 
of AINS in turbulent  ISM. 
We consider constant ISM density and isotropic 
Kolmogorov turbulence\footnote{%
Any other more complicated model of turbulence requires better knowledge
of the ISM structure especially on small scales. This information can be
obtained from radio pulsar scintillation observations 
(see for example \cite{sh98}). 
We plan to include this data in our future calculations.} 
(Kolmogorov spectrum is in good correspondence with most of observation 
of interstellar, i.e. intercloud, turbulence, see for example \cite{f90}).
Orientations of turbulent cells at all scales are assumed to be independent. 
Turbulent velocities at different scales relate to each other according to
the Kolmogorov law:
$$
        \frac{v_t(r_1)}{v_t(r_2)} = \left(\frac{r_1}{r_2}\right)^{1/3}\,.
$$

Observations show (see \cite{rss88} and references therein) 
that turbulent velocity at the scale of
$R_t \simeq 2\cdot10^{20}\textrm{~cm} \simeq 70$~pc
is about  $v_t \simeq 10$~km s$^{-1}$. The scale above is close to the
thickness of the gas disk of the Galaxy, and the velocity above ---
to the speed of sound in the ISM.  It corresponds to the largest
cell size possible and the fastest movements 
(otherwise turbulence will efficiently dissipate its energy in shocks).

As an AINS moves through the ISM it can capture matter inside
so called Bondi (or accretion, or gravitational capture) radius, 
$R_G=2GM/v^2$. Here $G$ is the gravitational constant, 
$M$ is the mass of NS, $v=\sqrt{v^2_{NS}+v^2_s\,}$, 
$v_{NS}$ is the spatial velocity of a NS relative
to ISM, $v_s$ is the speed of sound (we assume $v_s=10$ km s$^{-1}$,
$M=1.4\, M_{\odot}$ everywhere in the paper; $v_s$ can be dependent on the
luminosity of AINS (\cite{bwm95}), but here we neglect it planning to iclude
this dependence in our future calculations).

Accretion rate $\dot M$ at the conditions stated above is equal to
(\cite{hl39}, \cite{bh44}): 
$$
        \dot M = \pi R_G^2 n m_p v\,,
$$
here $n$ is the  number density of the ISM, $m_p$ is the mass of proton.
This accretion rate corresponds to luminosity:
$$
L=GM \dot M/R \sim 10^{32} n \, (v/10 \, {\rm km}\, {\rm s}^{-1})^{-3} \, 
{\rm erg} \, {\rm  s}^{-1}
$$
Based on population synthesis models (\cite{pea00b}) we can expect,
that on average AINS should have luminosities about $10^{29}$ erg s$^{-1}$.
Simple calculations show, that most of this energy will be emitted in
X-rays with a typical blackbody temperature about 0.1 keV
(if due to significant magnetic field accretion proceeds 
onto small polar caps then the temperature would be higher up to 1 keV).
Note, that Bondi rate is just the upper limit, in reality
due to heating (\cite{s70}, \cite{s71}, \cite{bwm95}) 
and magnetospheric effects (\cite{ta2001}) the accretion rate can be lower.

Due to the turbulence accreted matter carries non-zero angular momentum:
$$
        j_t = v_t(R_G) \cdot R_G = v_t(R_t) R_t^{-1/3} R_G^{4/3}\,.
$$
In this formula  it is considered that cells of the size $r=R_G$
are the most important, otherwise in the Eq.(\ref{eq:D-diff}) below it is
necessary to introduce factor $\alpha\neq 1$. 

 If $j_t$ is larger than the Keplerian value on the magnetosphere boundary
(i.e. on the Alfven radius $R_A = (\mu^2/2\dot M\sqrt{GM\,})^{2/7}$, here
$\mu$  is the magnetic moment of a NS) an accretion disk is formed around the
AINS. In the disk a part of the angular momentum is carried outwards,
and the NS accrets matter with 
Keplerian angular momentum $j_K = v_K(R_A) \cdot R_A$, $v_K$ -- Keplerian
velocity. 
This situation holds only for very low magnetic fields and low
spatial velocities of NSs, 
thus we do not take it into account in the present calculations.

 We consider the lowest spatial velocity of NSs\footnote{%
The lowest measured velocity is $\ga50$~km~s$^{-1}$ (\cite{ll94}),
please bear in mind that 
selection effects are very important here.} to be equal to 10 km s$^{-1}$. 
This value is of order of the speed of sound. 
Therefore lower spatial velocities can not
change the accretion rate significantly. 
Moreover such low values are not very probable due to
non-zero spatial velocities of NSs progenitors 
(see \cite{pea00a}, \cite{acc01} for limits onto the fraction of low
velocity INSs derived by different methods). 
For several cases below we consider $v\simeq v_{NS}$.

 During the time required to cross a turbulent cell of the size $R_G$,  
$\Delta t = 
R_G/v_{NS} \simeq 11.8 \, (v/10 \, \textrm{km~s}^{-1})^{-3}$~yrs,
the change in the angular momentum of a NS $J$ is given as:
$$
        \left| \vec{\Delta J} \right| = \dot M j_t \Delta t
     =\pi n m_p v_t(R_t)R_t^{-1/3}R_G^{13/3}
$$
$$
            \simeq 1.23\cdot 10^{39}\, {\rm g\, cm^2\,  s}^{-1} \times
$$
$$
\, n \left(\frac{v_t(R_t)}{10\, {\rm km \, s}^{-1} }\right)
 \left(\frac{R_t}{2\cdot 10^{20}\, {\rm cm}}\right)^{-1/3} 
\left(\frac{v}{10 \, {\rm km \, s}^{-1}}\right)^{-26/3}\,.
$$

\noindent
Accordingly, the change of the spin frequency reads:
$$
        \left| \vec{\Delta\omega} \right| = \left| \vec{\Delta J}\right|/I
$$
$$
\simeq 1.23\cdot10^{-6} \, {\rm s}^{-1} \times
$$
$$
\, n \left( \frac{v_t(R_t)}{10\, {\rm km \, s}^{-1} } \right)
 \left(\frac{R_t}{2\cdot 10^{20}\, {\rm cm}}\right)^{-1/3} 
\left( \frac{v}{10 \, {\rm km \, s}^{-1}}\right)^{-26/3}  I_{45}^{-1}\,,
$$
here  $I=I_{45} 10^{45}$~g~cm$^2$ --- moment of inertia of a NS.
Orientation of $\vec{\Delta\omega}$ is random, and it is isotropically
distributed on the sphere. The value of the frequency change is strongly
dependent on the spatial velocity of the NS: $\Delta\omega \sim v^{-26/3}$, 
so the maximum value for $v_{NS}=10$~km~s$^{-1}$ is 
$\Delta\omega_{\max} \simeq 6\cdot10^{-8} $~rad s$^{-1}$.

 On this view, it follows that it is possible to describe the
spin evolution of an AINS as random movement in 3-D space of angular
velocities $\vec\omega$.  
Since typical temporal and ``spatial'' scales $\Delta t$ 
and $\Delta\omega$ are
reasonably small ($\Delta t \ll t_{gal}
\simeq 10^{10}$~yrs, $\Delta\omega \ll \omega  \sim 10^{-1}\div
10^{-7}$~rad s$^{-1}$) we
consider the problem to be continuous. Therefore, we can use differential
equations valid for continuous processes. 

 In this case the spin evolution of an AINS in the space  $\vec\omega$ 
is described by the diffusion equation with the coefficient, namely:
\begin{equation}
        D = \frac{\alpha}{6} \frac{\Delta\omega^2}{\Delta t},
\label{eq:D-diff}
\end{equation}
here $\alpha$ --- coefficient which takes into account cells with sizes
 $r\neq R_G$ (everywhere in this paper we use $\alpha=1$).

 Besides random turbulent influence an AINS is spinning down due to magnetic
braking:
\begin{equation}
\frac{d\vec{\omega}}{dt} = 
\kappa_t \frac{\mu^2}{IR_{co}^3} 
\frac{\vec{\omega}}{|\omega|} + \vec F_t =
\kappa_t \frac{\mu^2}{IGM}         
\omega^2 \frac{\vec{\omega}}{|\omega|} + \vec F_t\,.
\label{eq:dwdt}
\end{equation} 
Here $R_{co} = (GM/\omega^2)^{1/3}$ --- corotation radius,
$\kappa_t$ --- constant of the order of unity, which takes
into account details of interaction between the magnetosphere and
accreted matter (everywhere in this paper we accept $\kappa_t=1$),
$\vec F_t$ --- random (turbulent) force with zero average value:
$\langle \vec F_t\rangle =0$. 

On large time scale and for relatively short spin periods we can neglect
momentum of the accreted matter and initial period of NS $p_0$.
In that case solution of Eq. (\ref{eq:dwdt}) looks as:
$$
        \omega \propto t^{-1}\,, \quad p \equiv \frac{2\pi}{\omega}
                \propto t\,.
$$

Magnetic braking produces a convective term
in the evolutionary equation for spin frequency
distribution,  $f(\omega)$.
As far as initial distribution of spin vectors is isotropic
(magnetic braking does not change the orientation of $\vec\omega$)
and turbulent diffusion is isotropic too, we obtain spherically symmetric
distribution in the space of angular velocities, i.e.:
$$
        f(\vec\omega) = f_3(\omega)
$$
($f_3(\omega)$ describes 3-D distribution of AINSs, its dimension is
$[\textrm{s}^{-3}]$).

 From (\ref{eq:D-diff}) and (\ref{eq:dwdt}) we obtain the evolutionary
equation:\begin{equation}
        \frac{\partial f_3}{\partial t} =
        \frac{A}{\omega^2} \frac{\partial}{\partial\omega}
        \left( \omega^4 f_3\right)
        + \frac{D}{\omega^2} \frac{\partial}{\partial\omega}
        \left( \omega^3 \frac{\partial f_3}{\partial\omega}\right)\,,
\label{eq:f3} 
\end{equation}
here $A=\left(\mu^2/IGM\right)$.

 Boundary condition at  $\omega=0$ can be derived from the equality of the
flow of the particles at this point to zero:$$
        \left.\frac{\partial f_3}{\partial\omega}\right|_{\omega=0} = 0\,.
$$

 It is easy to find stationary solution of the equation (\ref{eq:f3}) on the
semi-infinity axis\footnote{%
As far as NS enter the stage of accretion with relatively short periods
$\omega_{\mathrm{\scriptsize A}}\gg\langle\omega(f_3^\textrm{\scriptsize \,st})\rangle$,
the obtained distribution is differed from the real one only on high
frequencies $\omega\simeq\omega_{\mathrm{\scriptsize A}}$.}
($\omega\geq0$)
\begin{equation}
        f_3^\textrm{\scriptsize \,st} = 
C_{st}\exp\left(-\,\frac{\mu^2}{3\,IGMD}|\omega|^3\right)\,,
\label{eq:f3_stac}
\end{equation}
here normalization constant $C_{st}$  
can be derived from the condition $4\pi\int
f_3^\textrm{\scriptsize \,st}(\omega)\omega^2 d\omega = N$,
$N$ --- full number of AINSs in the distribution. Total number of NSs
in the Galaxy is uncertain, $N\sim 10^8$--$10^9$. Population synthesis
calculations (\cite{pea00b}) give arguments for higher total number about
$10^9$. Here we do not address this question. From the curves below, which
represent {\it relative} period distribution of AINSs,
one can  determine absolute numbers of AINSs with each period value
if the total number of these objects is known. 

Position of the maximum of $f_3^\textrm{\scriptsize \,st}(p)$ 
depends on $v$, $\mu$, $n$ and
$M$. Increasing of $v$ and $\mu$ shifts the maximum to larger $p$,
increasing of $n$ and $M$ --- to shorter $p$.

 If we have to solve the problem for constant starformation rate
we can write second boundary condition as:
$$
        f_3(\omega_{\mathrm{\scriptsize A}}) = \frac{IGM}{4\pi\kappa_t\mu^2}
                \cdot \frac{\dot N}{\omega_{\mathrm{\scriptsize A}}^4}\,,
$$
here $\dot N$ --- rate at which NSs come to the stage of accretion,
$\omega_{\mathrm{\scriptsize A}} = 2 \pi/p_A$ --- 
{\it accretor} frequency, at that
value for given $v$ and $n$ accretion sets on,
$p_A=2^{5/14}\pi(GM)^{-5/7}(\mu^2/\dot M)^{3/7}
\simeq 300\, \mu_{30}^{6/7} (v/10\, {\rm km}\, {\rm s}^{-1})^{9/7}
n^{-3/7}\, {\rm s}$.  

 The function $f_3(\omega)$ is connected with distributions of
absolute values of the angular velocity, $f_1(\omega)$,
and of the spin period,  $f(p)$, according to the formulae:
$$
        f_1(\omega) = 4\pi\omega^2 f_3(\omega)\,,
$$
$$
        f(p) = \frac{32\pi}{p^4} f_3\left(\frac{2\pi}{p}\right)\,.
$$

  Figure 1 shows the evolution of $f(p)$ 
as a function of time for typical parameters
of an AINS and the ISM.

\begin{figure}
\vbox{\psfig{figure=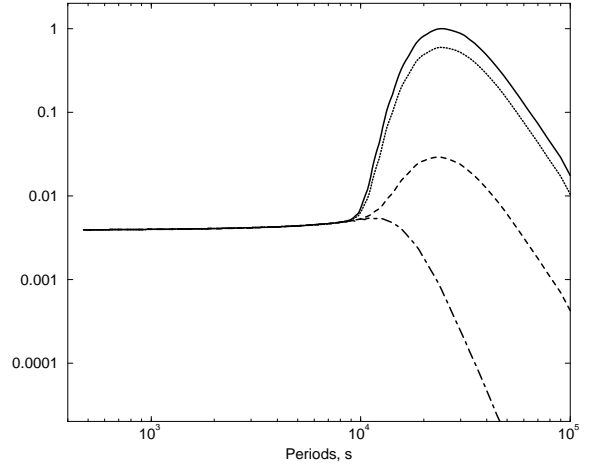,width=9.0cm,angle=-90}}
\caption[]{Evolution of the period distribution in time; 
$\mu=10^{30}$~G~cm$^{3}$, $n=1$ cm$^{-3}$, $v_{NS}=10$~km~s$^{-1}$. 
Curves are plotted for
4 different moments from $1.72\cdot 10^9$ yrs to $9.8\cdot 10^9$ yrs
($t_A$ for chosen parameters is equal to $\simeq 1.7\cdot 10^9$ yrs). 
An AINS crosses the horizontal part from $\simeq 10^2$~s to $10^4$~s
in $\sim 6\cdot 10^7$~yrs.
Curves were normalized to 1 in the maximux for the highest curve.
}
\end{figure}

For given $v$, $\mu$ and $n$ NSs enter the {\it accretor} stage when
they spin down to  $p_{\mathrm{\scriptsize A}}$. Up to this value they evolve
as {\it ejectors} and {\it propellers} (\cite{li92}, \cite{trento2000}).

 For $p_{\mathrm{\scriptsize A}}<p<p_{cr}$
 influence of
the turbulence is small and AINS spin down according to  $p \propto t$.
Here $p_{cr}$ defines the stage when spin changes due to magnetic braking
and turbulent acceleration/decelaration become of the same order of
magnitude (see below the Discussion).
One can  introduce another useful timescal, $\Delta t_{cr}$. 
This is the time required to reach turbulent regime, $p=p_{cr}$.  
These considerations
together with constant starformation rate  determine
left part of the distribution where $f(p) = const$ (see Fig.~1).

 As the period grows turbulence becomes more and more important.
Finally the first AINSs reach the point $p=\infty$ ($\omega=0$)
and there  equilibrium component of the distribution 
(described by Eq. (\ref{eq:f3_stac}) is formed quickly.
This component decreases in a power law fashion 
($f(p) \propto p^{-4}$) at $p\gg p_{turb}$ (where $p_{turb}$ is determined
by the width of the distribution (\ref{eq:f3_stac})), 
and decreases even 
faster than exponentially at $p\ll p_{turb}$.
Please note that the number of AINSs reaching equilibrium grows linearly
with time. As a result, the amplitude of equilibrium component of the
distribution grows likewise.
If the time required to reach the {\it accretor} stage, $t_A$,
is large ($t_{gal}-t_{\mathrm{\scriptsize A}}\ll t_{gal}$) 
the equilibrium component is not formed. 
For $t_{\mathrm{\scriptsize A}}>t_{gal}$ 
NS never reach the stage of accretion.

 Similar considerations can be applied also to binary X-ray pulsar
(see \cite{li92}). In binaries NSs can reach real period equilibrium
(\cite{gl79}), and stationary solution should be used.
In that case observations of period
fluctuations can help to derive physical parameters of the systems,
for example stellar wind velocity for wind-fed pulsars (\cite{lp95b}).

\section{Calculations and results}

 The main aim of this paper is to obtain period distribution for AINS
in turbulized ISM. We assume that NS are born with short ($\ll 1$~s)
spin periods. Magnetic moment distribution is taken in log-Gaussian form:
\begin{equation}
        f(\mu) = \frac{1}{\sqrt{2\pi\sigma_m\,}}
                \exp\left(\frac{(\log\mu -
        \log\mu_0)^2}{2\sigma_m^2}\right)\,,
\label{eq:f(mu)}
\end{equation}
with $\log\mu_0 = 30.06$ and $\sigma_m=0.32$ 
(see \cite{trento2000} 
and references therein for a recent discussion and data 
on magnetism in NSs).
Magnetic field is  considered to be constant (see,
for example, \cite{ugk96} for calculations of magnetic field decay
and related discussion).

 Velocity distribution (due to kick after supernova explosion)
is assumed to be Maxwellian:
\begin{equation}
 f(v_{NS})=
\frac{6}{\sqrt{\pi}}\frac{v_{NS}^2}{v_m^2}
\exp\left(\frac{3}{2}\frac{v_{NS}^2}{v_m^2}\right)
\label{eq:f(v)}
\end{equation}
 with $v_m=200$~km~s$^{-1}$  
which corresponds to $\sigma_v\simeq 140$~km~s$^{-1}$
(see for example \cite{ll94}, \cite{cc97},  
\cite{hp97}, \cite{letal97}, \cite{cc98}, and 
\cite{acc01} for discussions on the kick velocity of NSs).

 We use two values of the ISM number density, $n=1$ cm$^{-3}$ 
and $n=0.1$ cm$^{-3}$, as typical values for the Galactic disk
\footnote{%
This assumption is reasonable for relatively low velocity INS,
but only this fraction of the whole population is important for us here.
}.

 All NS are divided into 30 groups in $\mu$ interval from $10^{28.6}$ to
$10^{31.6}$ G$\cdot$cm$^{3}$ with the step 0.1~dec 
and 49 groups in velocities from 10 to 500~km~s$^{-1}$ with the step
10~km~s$^{-1}$.

 For each group we calculate $t_A$, the time necessary to reach accretion:
$$
        t_A = t_E + t_P\,,
$$
here $t_E$ and $t_P$ --- 
durations of the {\it ejector} and {\it propeller} stages
correspondently.
 
 On the stage of ejection a NS spin down due to magneto-dipole losses:
$$
        t_E \simeq 10^9 n^{-1/2} 
        \left( \frac{v}{10~\textrm{km}\, \textrm{s}^{-1}} \right)
        \left( \frac{\mu}{10^{30}~\textrm{G}\cdot\textrm{cm}^3}\right)^{-1}
        ~\mbox{yrs.}
$$

 Spin down on the {\it propeller} stage (\cite{s70b}, \cite{is75}) 
is not well known, but for constant field
always $t_P<t_E$ (\cite{lp95}) (see also results of numerical calculations
in \cite{t99}), 
so we neglect $t_P$ in the rest of the paper,
i.e.  we assume $t_A=t_E$.

 For high spatial velocities after {\it ejector} an INS appears not as a
{\it propeller}, but as a so-called {\it georotator} (\cite{li92}, see also
\cite{r2000}). At this stage $R_A>R_G$, so surrounding matter does not feel
gravitation of the NS, and its magnetosphere looks like magnetospheres
of the Earth, Jupiter etc. Partly because of that effect  we truncated
our velocity distribution at 500~km~s$^{-1}$.

 For systems with $t_A<t_{gal} = 10^{10}$~yrs
we solve Eq. (\ref{eq:f3}) on the time interval $t_{gal}-t_A$.
For angular velocities we use grid with 200 cells from 
$\omega=0$ to $\omega_A$. Conservative implicit scheme is used.
Derived distributions $f(p,\mu,v)$ (for different groups)  
are summed with weights from Eqs. (\ref{eq:f(mu)}) and (\ref{eq:f(v)}).

\begin{figure}
\vbox{\psfig{figure=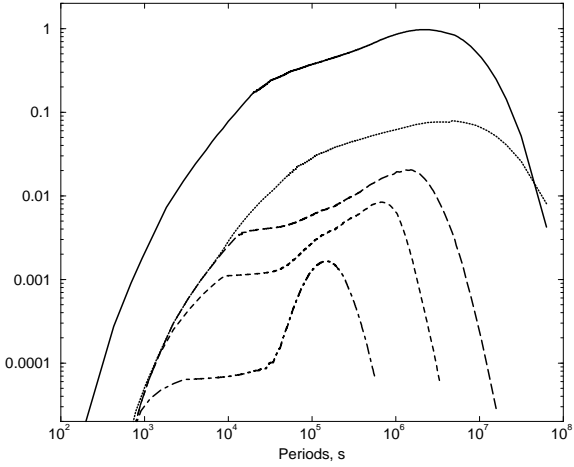,width=9.0cm,angle=-90}}
\caption[]{Period distribution for populations of AINSs evolving in the ISM
with number density $n=1$ cm$^{-3}$ (upper solid curve) and $n=0.1$
cm$^{-3}$
(the second dotted curve). Lower curves show results for low velocity
part of AINS population for $n=0.1$ cm$^{-3}$ 
($v<60,\, 30,\, 15$~km~s$^{-1}$ correspondently, see Discussion below). 
Curves for normalized to 1 in maximum for the highest curve.}
\end{figure} 

 Final distribution for $n=1$ cm$^{-3}$ and $n=0.1$ cm$^{-3}$
are shown on Fig.~2.

 We note, that the final distribution significantly differs from the one
for a single set of parameters ($n, \mu, v$): compare Fig.~1 and Fig.~2.
In the final one we see power-law cutoff ($f \propto p^{-4}$)
at long periods ($p>10^7$~s) similar to cutoff on the Fig.~1.
Then at the intermediate periods ($10^4<p<10^7$~s) the curve has
power-law ascent due to summation of individual curves maxima.
Finally a sharp cutoff at short periods  ($p<10^4$~s) is presented
because different INSs enter the {\it accretor} stage with different 
periods: $p_A=p_A(n, \mu, v)$.

\section{Discussion and conclusions}

 After an INS comes to the stage of accretion ($p>p_A$) its spin 
is controlled by two processes (see Eq.2): 
magnetic spin-down and turbulent spin-up/spin-down. Schematic view 
of such evolution process is shown on Fig.~3.

\begin{figure}
\vbox{\psfig{figure=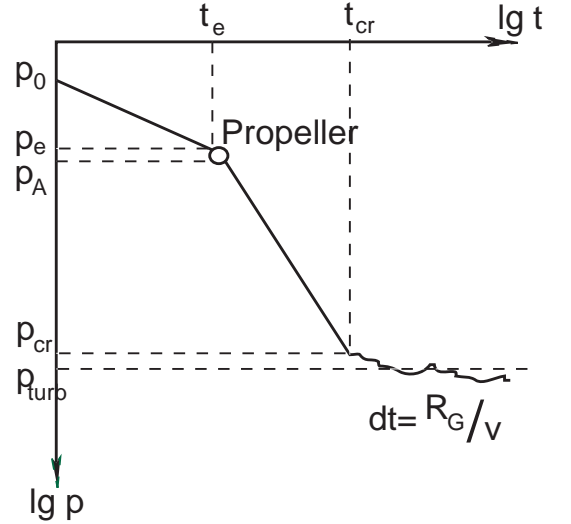,width=9.0cm}}
\caption[]{Schematic view of period evolution of an INS.
A NS starts with $p=p_0$, then spins down according to magneto-dipole
formula up to $p_E$, then a short {\it propeller} stage 
(marked by a circle) appears and lasts
down to $p=p_A$. 
At the situation illustrated in the figure $t_E\approx t_A$.

On the stage of accretion at first magnetic braking is more
important, after at $t=t_{cr}$ turbulence starts to influence significantly
spin evolution of a NS period stars to fluctuate with a typical time scale
$R_G/v_{NS}$ close to mean value noted $p_{turb}$ on the vertical axis.}
\end{figure} 

We can describe them with characteristic timescales: $t_{mag}$ and
$t_{turb}$.
As one can calculate, $t_{mag} \sim p$ and $t_{turb} \sim p^{-1}$.
Here we evaluate $t$ as $p/\dot p$, and
$\dot p_{turb}= p^2 \dot M j / (2 \pi I)$.
Here we again note, that $j_t$ can be larger that the Keplerian value
of angular momentum
on the magnetosphere, $j_K=v_K(R_A), R_A$. So we write $j$ instead of
$j_t$ as we did before, now $j={\min}(j_K, j_t)$.

\begin{equation}
\dot p_{mag}=2 \pi \mu^2/(IGM)
\end{equation}

Initially (immediately after $p=p_A$) magnetic spin-down is more
significant:

\begin{equation}
t_{mag}=\frac{IGM}{2\pi \mu^2}p,
\end{equation}
but at some period, $p_{cr}$, these two timescales should become
of the same order, and for longer periods an INS will be governed mainly by
turbulent forces.

 One can obtain the following formula:

\begin{equation}
 p_{cr}^2=\frac{4 \pi^2 \mu^2}{GM\dot M j}.
\end{equation}

This period lies between $p_A$ and $p_{turb}$.
An INS reach $p_{cr}$ in $\Delta t_{cr}\sim 10^5$--$10^7$ yrs 
after onset of accretion:

\begin{equation}
 \Delta t_{cr}= \frac{I \sqrt{GM}}{\mu \sqrt{\dot M j}}.
\end{equation}
Here $\Delta t_{cr}$ are calculated approximately as $p_{cr}/\dot p_{mag}$.

 The period evolution of an AINS can be described in the  
following way (Fig.~3): 
after the INS comes to the
stage of accretion it spins down for $\sim 10^5-10^7$ yrs up to $p_{cr}$,
then the evolution is mainly
controlled by the turbulence, and period fluctuates
with typical value $p_{turb}$, which is determined by the properties of the
surrounding ISM and spatial velocity of an INS.

We note, that we do not take into account any selection effects.
For example, as far as period and luminosity both depend on the
velocity of an INS, they are correlated. Taking into account
flux limits of the present day satellites it is possible to
calculate probability to {\it observe} an accreting INS with
some period.
 Also periods  (and luminosities)
are correlated with position of an INS in the Galaxy.

So, it is reasonable to make calculations which include all that effects
in order to make better predictions for observations. We plan to unite
our population synthesis calculations with detailed calculations of
spin evolution later.
In this paper we present distributions {\it as if} all accreting INSs
can be observed.

 For the main part of its life spin period of a low-velocity AINS
is governed by turbulent forces. Characteristic timescale for them can be
written as: $t_{turb}=I\omega/(\dot M j)$. And it is equal to
$10^4$-$10^5$ yrs for typical parameters.

 For field decay picture should be completely different (see for example
\cite{kp97}, \cite{w97}),
and observations of AINSs can put important limits onto models
of magnetic field decay in NS (\cite{pp2000}). 
For decaying fields AINSs can appear as pulsating sources with
periods about 10 s and $\dot p$ about $10^{-13}$ s/s (\cite{pk1998}).
The value and sign of $\dot p$ will fluctuate as an INS
passes through the turbulent cell on a time scale 
$R_G/v_{NS}$, which is about a year for typical parameters.
Irregular fluctuations of $\dot p$ on that time scale can be significant
indications for accretion (vs. cooling) nature of observed luminosity.

 Roughly $\dot p$ can be estimated from the expression:
\begin{equation}
|\dot p|\le p^2\frac{\dot M j}{2\pi I}\sim v^{-17/3}.
\end{equation}
Here we neglect magnetic braking in Eq. (2).
This equation for $\dot p$ 
is valid for turbulent regime of spin evolution, 
i.e. for $p\gg p_{turb}$.
For the given $p$ 
value of $\dot p$ fluctuates between $+p^2\frac{\dot M j}{2\pi I}$ and
$-\,p^2\frac{\dot M j}{2\pi I}$, and depends on $n$, $v$, but not on
$\mu$ (if $j=j_t$, not $j=j_A$). In this picture $\dot p$ distribution
in the interval specified above (Eq.11) is flat.

 Behavior of $p$ and $\dot p$ of AINSs in molecular clouds
can be different (\cite{campana93}), 
especially for low spatial velocities of NSs.
Some of our assumptions in that case are not valid, and results
cannot be applied directly. But we note, that passages through 
molecular clouds are relatively rare and short, they cannot significantly
influnce the general picture of AINSs spin evolution.

 Period distribution which can be obtained from observation 
(for example from \emph{ROSAT} data) can be different
from two upper curves on the Fig.~2. 
Such surveys are flux-limited, so they include
(in the case of AINS) the most luminous objects. But they form only a small
fraction of the whole population. 

 To illustrate it on the Fig.~2 we also plot distributions
for low velocity objects ($v<60, 30, 15$~km~s$^{-1}$).
In the case of fixed ISM density an upper limit on the value of the spatial
velocity corresponds to a lower limit on the accreting luminosity of an AINS.
Clearly, brighter the source --- shorter (on average) its spin period.
Even in that groups with relatively short periods their values are far
from typical periods of \emph{ROSAT} INSs, $\sim 5-20$ s.
So, these objects cannot be explained by {\it accretors} with constant field
$B\sim 10^{12}$ G.
 
 Calculations of period distributions for decaying magnetic field,
for populations of INSs with significant fraction of magnetars and for
accretion rate different from the standart Bondi-Hiyle-Littleton value
(due to heating and influence of a magnetosphere) will be done in a separate
paper.

 In conclusion we stress readers attention on the main results of the paper:

\begin{itemize}
\item[---]
we obtained
spin period distributions for AINSs with constant magnetic field
and ``pulsar'' properties (field, initial periods and velocity
distributions).

\item[---]
these distributions are shown in Fig.~2. They have 
broad maximum at very long periods, $\sim 10^6$ s.
In that case observed objects should not show any periodicity.

\item[---]
periods of these objects should fluctuate on a time scale $R_G/v_{NS}\sim 1$ yr.
\end{itemize}

\begin{acknowledgements}
This work was supported by grants of the 
RFBR 01-02-06265, 00-02-17164, 01-15(02)-99310.

AK thanks Sternberg Astronomical Institute for hospitality.
SP and MP thank Monica Colpi, Roberto Turolla 
and Aldo Treves for discussions and
Universities of Milano and Padova for hospitality.

We thank Vasily Belokurov and the referee of the paper
for their comments on the text and useful suggestions.
\end{acknowledgements}

\end{document}